\documentclass[12pt,preprint]{aastex}

\newcommand{\smc}{CXOU~J010043.1$-$721134}

\begin{document}


\title{A search for the optical counterpart to the magnetar \smc.}


\author{M. Durant}
\affil{Instituto de Astrof\'isica de Canarias, La Laguna, Tenerife
    Spain}
\email{durant@iac.es}

\and

\author{M. H. van Kerkwijk}
\affil{University of Toronto, Toronto, Canada}


\begin{abstract}
After our tentative detection of an optical counterpart to \smc\ from
archival Hubble Space Telescope (HST) imaging, we have followed up
with further images in four bands. Unfortunately, the source
originally identified is not confirmed. We provide deep photometric limits
in four bands and accurate photometry of field stars around the
location of the magnetar.
\end{abstract}


\section{Introduction}
Magnetars are neutron stars that derive their large X-ray luminosity
from the decay of a super-strong magnetic field or the order
$10^{15}$\,G (and even greater internally), many orders greater even
than normal radio pulsars and accreting X-ray sources (Woods \&
Thompson, 2006). They are found
preferentially in the Galactic plane, in accordance with their
presumed youth and high-mass progenitors (e.g., Figer et al., 2005). This makes
observations in soft X-rays and the optical difficult, due to the
large columns of extincting material to each of the objects (Durant \&
van Kerkwijk, 2006a).

For the nearest of the magnetars, 4U 0142+61, an intriguing
break was seen in the broad-band optical spectrum between the B and U
bands (Hulleman et al., 2004). Due to the faintness and extinction to
this source, it has not proved possible so far to further characterize
the optical feature.

A source for which the extinction will be much less of an issue is
\smc, a magnetar in the Small Magellanic Cloud.
This source was detected by {\em Chandra} as a slow-spinning X-ray source,
with a bright thermal or power-law spectrum (Lamb et al., 2002).
Although relatively faint in the first observations by Lamb et al. and
Majid et al., in later {\em Chandra} and {\em XMM} observations, the
object had brightened somewhat towards a similar luminosity to the
other Anomalous X-ray Pulsars ($\sim2\times10^{35}$\,erg\,s$^{-1}$,
0.5--10\,keV range; McGarry et al., 2005). The AXPs (the more stable
type of magnetar) all seem to have the same 2--10\,keV luminosity
($\sim 1\times 10^{35}$\,erg\,s$^{-1}$; Durant \& van Kerkwijk,
2006b).

In Durant \& van Kerkwijk (2005a, hereafter DvK05), we presented
evidence for the detection of an optical counterpart to \object{CXOU
  J010043.1-721134} in archival Hubble Space Telescope imaging with the Wide
Field/Planetary Camera 2 (WFPC2). The serendipitous detection was
based on a single exposure from a survey of the SMC
(\dataset[MAST,mission=hst&dataid=U6744102R]{Tolstoy, 1999}). Although a faint source, the detection
parameters and statistics of non-detections in the field pointed to it
likely being a true detection.

Here we present deeper HST imaging of the field of \smc\ in four bands to
attempt to confirm the tentative detection presented in DvK05.

\section{Observations}
Following the failure of the Advanced Camera for Surveys (ACS), we
planned an imaging campaign with the Wide Field/Planetary Camera 2
(WFPC2) on board the Hubble Space Telescope (HST), aiming first to
confirm the detection in DvK05, and second for broad-band photometry
to measure the spectral energy distribution throughout the optical and
into the ultra-violet. We used the filters F336W, F439W, F606W and
F814W and the wide-field chips, for reduced read-out noise across the
PSF area. Exposures were dithered by non-integer pixel shifts, in order
to sample the PSF and decrease the sensitivity to bad pixels.
We obtained 6 exposures in F336W (total time: 8400\,s), 6 exposures in
F439W (3400\,s), 2 exposures in F606W (800\,s) and 4 exposures in
F814W (2600\,s).

The images were analyzed with the {\tt HSTphot} package (Dolphin,
2000), which handles bad pixel rejection, sky level estimation, cosmic
ray identification, image alignment, PSF analysis and photometry for
the ensemble image set. 

\begin{figure}
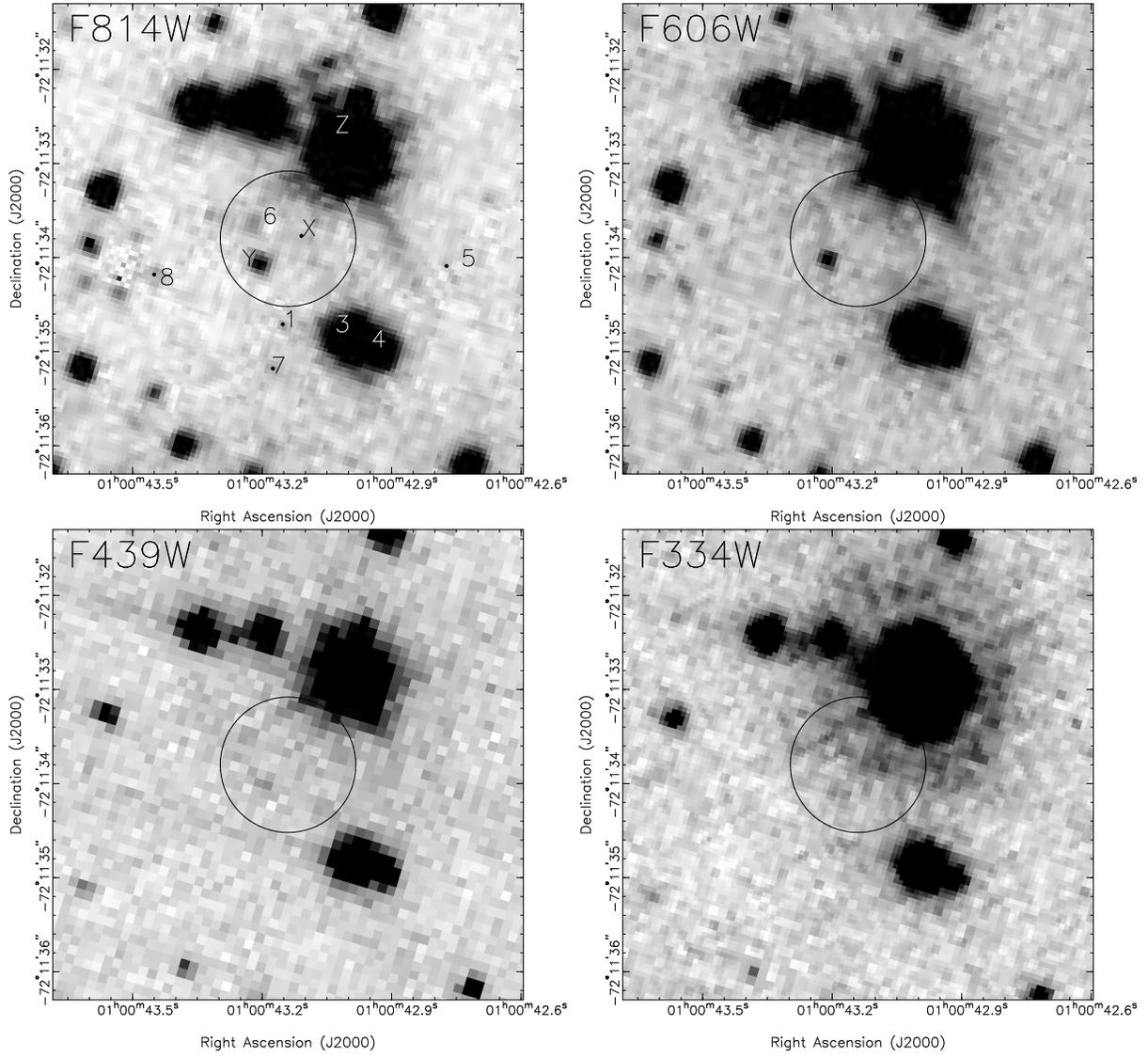

\begin{center}
\includegraphics[angle=0,width=0.48\hsize]{f1a.eps}
\includegraphics[angle=0,width=0.48\hsize]{f1b.eps}\\
\includegraphics[angle=0,width=0.48\hsize]{f1c.eps}
\includegraphics[angle=0,width=0.48\hsize]{f1d.eps}
\caption{Images of the field of \smc\ in the F814W, F606W, F439W and
  F334W filters (left to right, top to bottom). The
  positional 90\% confidence circle based on the {\it Chandra} detection
  is shown, and the stars listed in Table~\ref{phottable} are
  labelled.}\label{images} 
\end{center}
\end{figure}

To stack together images (for display and astrometry
only), we over-sampled each frame by a factor of two, and aligned and
added these, removing cosmic rays in the process.  See
Figure~\ref{images} for co-added 
images in each band. Clearly the source named Star X DvK05 is not
present at the same location. 

To register the images to the ICRS, we searched the astrometric
catalogs for references in the field. Both the
USNO B1.0 (Monet et al. 2003) and GSC II (STScI, 2006) have substantial scatter in star
locations compared to the images and to each other in this crowded
field. The UCAC catalog (Zacharias et al., 2003), generally regarded as having more
precise positions, has too few stars in the field for an astrometric
solution. We used the 2MASS (Cutri et al., 2003) catalog, primarily on the F439W
image (in which the 2MASS stars were not saturated). 11 2MASS stars
were matched on the WF3 chip, of which 9 were usable to fit for
rotation, scale and offset. Final residuals were $\sim0.08$\arcsec in
each coordinate. The systematic uncertainty of the 2MASS positions is
estimated to be 0.1--0.2\arcsec; if we assume the worst case, then the
0.6\arcsec {\em Chandra} positional uncertainty (90\% confidence)
translates into a 0.72\arcsec positional uncertainty (90\% confidence)
on our images. Transferring the astrometric solution from the F439W
image to the other images incurred negligible additional uncertainty. 

Table~\ref{phottable} lists the positions and magnitudes of stars
around the inferred X-ray position of \smc. Comparing to the images in
Figure~\ref{images}, one can see that some of these sources are not
good detections.   

The $3\sigma$ photometry limits for the field were: 24.2 in F334W,
25.6 in F439W, 26.2 in F606W and 25.9 in F814W. These magnitudes are
{\em flight system}: defined such that a star of zero color in the
Johnson-Cousins UBVRI system has zero color in any pair of WFPC2
filters, and the F555W magnitudes $m_{555}=V$ (Holtzman et
al. 1995). A color-magnitude 
diagram of stars in the field is shown in Figure \ref{CMD} for the two
filters F606W and F814W, which have the highest number of
detections. All of the detected stars are consistent with being part
of the normal stellar population in the field.

\section{Discussion}
\begin{figure}
\begin{center}
\includegraphics[angle=0,width=0.7\hsize]{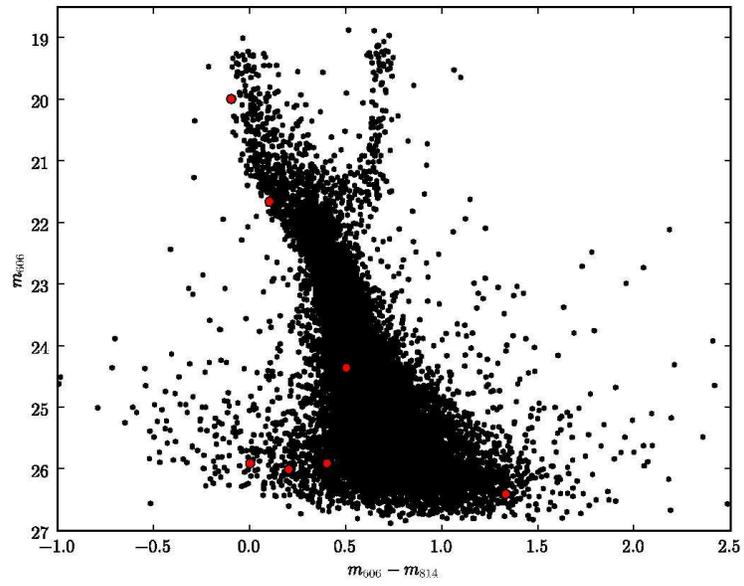}
\caption{Colour-magnitude diagram of stars in the field of \smc, for
  the filters F606W and F814W. Circular markers represent the few
  sources near to or in the positional error circle with measured
  magnitudes. }\label{CMD} 
\end{center}
\end{figure}

Given the non-detection probability of $\sim1.5$\% estimated in DvK05,
it is rather surprising that the deeper 
observations presented here failed to detect the same source. The
estimate was based on the number of good detections in the field which
would also have been considered interesting (by color), had they fallen within
the positional error circle. Only one new source is detected within
the positional error circle, Star 6, and this is red enough in
$m_{606}-m_{814}$ to appear as a normal star on the color-magnitude
diagram, Figure \ref{CMD}

Star Y  does not appear to have varied significantly
between the two observations in either filter where it was detected,
and Star Z  is still consistent with a bright early-type
star. 

Two distinct possibilities exist for the non-detection of the
counterpart to \smc\ proposed in DvK05: the original detection was
false, or the detection was real, but the source has faded
considerably in the intervening time.
For the detected object to physically leave the positional uncertainty circle at
the distance of the SMC, or even by a foreground white dwarf is not
possible, the proper motion required would be too great.

One obvious solution is that the detection in DvK05 was, in fact, a
cosmic ray or some other artifact. {\tt HSTphot} rejects detections as
cosmic rays if they are much sharper than typical stars, and this
source was, if anything, more diffuse. Some objects were, however, in
the same part of the color-magnitude diagram as Star X, so there is a
small but non-negligible chance that this was a mere fluke. In this deeper
set of observations, there remain a few detections of similar
brightness in $m_{606}$ and blueness in $m_{606}-m_{814}$ as the
original detection. The large scatter at the bottom of Figure
\ref{CMD} is not significant in this respect, as these sources
(including the couple near the error circle) are more poorly measured
than Star X appeared to be in DvK05.

Although the detection in DvK05 is clearly in doubt following these
observations, it is possible that the object faded considerably in the
optical, and became undetectable. The only AXP that has been detected
multiple times in the optical, 4U 0142+61, does at times show large
variations in flux, and the X-ray to optical flux ratio inferred in DvK05 was
much larger for Star X than it had typically been for 4U
0142+61. Assuming that none of the sources detected is the counterpart
to \smc, we derive a limit on the flux ratio $f_X/f_V>114$ (with
$V\approx m_{606}$, $A_V=0.3$ from Hilditch et al., 2005; and X-ray flux in the
2--10\,keV range from Woods \& Thompson, 2006), which
compares to a typical value of 460 for 4U 0142+61. Thus, the general
consistency found between different AXPs (Durant \& van Kerkwijk,
2005b) could still hold for this source as well.


To summarize, follow-up HST/WFPC2 observations of the field of
\smc\ have failed to confirm our earlier tentative detection of an
optical counterpart. No convincing counterpart is seen, with much
better limiting magnitudes than before. The absence of a detection
could either mean that the original detection was false, or that the
counterpart has faded significantly. If the latter, its X-ray to
optical flux ratio could now be the same as for 4U 0142+61.

\acknowledgments
MD is funded by the Spanish Ministry of Science. MHvK acknowledges the
support of NSERC.  

{\it Facilities:} \facility{HST (WFPC2)}.

\clearpage

\begin{deluxetable}{lllllll}
\tabletypesize{\scriptsize}
\tablecaption{Photometry of field stars\label{phottable}}
\tablewidth{0pt}
\tablehead{
\colhead{Label} & \colhead{RA} & \colhead{dec} & \colhead{$m_{334}$} &
\colhead{$m_{439}$} & \colhead{$m_{606}$} & \colhead{$m_{814}$} }
\startdata
1&1:00:43.152&-72:11:34.71& 24.2$\pm$0.3& $>$25.6&26.2$\pm$0.3& $>$25.9\\
Y&1:00:43.209&-72:11:34.09& 24.1$\pm$0.3& 24.81$\pm$0.15& 24.35$\pm$0.08& 23.85$\pm$0.06\\
3&1:00:42.995&-72:11:34.92& 19.031$\pm$0.005& 19.966$\pm$0.006& 19.986$\pm$0.007& 20.084$\pm$0.005\\
4&1:00:42.926&-72:11:35.03& 21.73$\pm$0.03& 21.858$\pm$0.017& 21.653$\pm$0.011& 21.551$\pm$0.012\\
5&1:00:42.773&-72:11:34.09&$>$24.2&$>$25.6&26.4$\pm$0.3&$>$25.9\\
6&1:00:43.221&-72:11:33.64&$>$24.2& $>$25.6& 25.9$\pm$0.2& 25.07$\pm$0.14\\
10&1:00:43.175&-72:11:35.18& $>$24.2& $>$25.6& 25.9$\pm$0.2& 25.8$\pm$0.3\\
Z&1:00:43.001&-72:11:32.94&\nodata&17.783$\pm$0.002&\nodata&\nodata\\
12 &1:00:43.450&-72:11:34.18& $>$24.2 & $>$25.6 & 26.1$\pm$0.2& 25.9$\pm$0.3
\enddata
\tablecomments{See Figure \ref{images} for the locations of the stars in the
  field. Uncertainties are formal 1$\sigma$ errors, limits are 95\%
  confidence.  Magnitudes are Flight-System (Dolphin, 2000), some are
  clearly spurious (see text).}
\end{deluxetable}


\begin{thebibliography}{}
\bibitem{a}
Cutri, R., Skrutskie, M., Van Dyk, S., Beichman, C., Carpenter, J.,
Chester, T., Cambresi, L., Evans, T., el al., 2003, Vizier catalog II/246
\bibitem{fish}
 Dolphin, A., 2000, PASP, 112, 1383
\bibitem{b} Durant, M. \& van Kerkwijk, M., 2005a, ApJ, 628, L135
\bibitem{gratuitous}
Durant, M., \& van Kekwijk, 2005b, ApJ, 627, 376 
\bibitem{me}
Durant, M. \& van Kerkwijk, M., 2006a, ApJ, 650, 1070
\bibitem{andagain}
Durant, M. \& van Kerkwijk, M., 200ba, ApJ, 650, 1082
\bibitem{hil}
Hilditch, R., Howarth, I., Harries, T., 2005, MNRAS, 357,304
\bibitem{hst}
Holtzman, J., Burrows, C., Casertano, S., Hester, J., Trauger, J.,
Watson, A., Worthey, G., 1995, PASP, 107, 1065
\bibitem{H04}
 Hulleman, F., van Kerkwijk, M., \& Kulkarni, S., 2004, A\&A, 416, 1037
\bibitem{gsc}
Guide Star Catatog 2.2, Space Telescope Science Institute (STScI) \&
Osservatorio Astronomico di Torino, 2006, Vizier catalog I/305
\bibitem{c} Figer, D., Najarro, F., Geballe, T., Blum, R., Kudritzki,
  R., 2005, ApJ, 622, L49
\bibitem{sheep}
 Lamb, R, Fox, D., Macomb, D., Prince, T., 2002, ApJ, 574, L29
\bibitem{coords}
 Majid, W., Lamb, R., Macomb, D., 2004, ApJ, 609, 133
\bibitem{mc}
McGarry, M., Gaensler, B., Ransom, S., Kaspi, V., Veljkovik, S., 2005, ApJ
627, L137
\bibitem{d} Monet, D., Levine, S., Canzian, B., Ables, H., Bird, A.,
  Dahn, C., et al., 2003, AJ, 125, 984
\bibitem{e} Zacharias, N., Urban, S., Zacharias, M., Wycoff, G., Hall,
  D., Monet, D., Rafferty, T., 2004, AJ, 127, 3043
\bibitem{PI}
 Tolstoy, E., 1999, IAU Symposium 192, ASP, eds Whitelock, P. and
 Cannon, R.
\bibitem{wood}
 Woods, P., \& Thompson, C., 2006, in ``Compact stellar X-ray sources'',
 eds Lewin, W., van der Klis, M., Cambridge University Press
\end{thebibliography}
\end{document}